\documentclass[twocolumn,showpacs,preprintnumbers,amsmath,amssymb,aps]{revtex4}
\usepackage{graphicx}

\newcommand{\up}                {\left|\uparrow\right>}
\newcommand{\down}          {\left|\downarrow\right>}
\newcommand{\downdown}{\left|\downarrow\downarrow\right>}
\newcommand{\downup}     {\left|\downarrow\uparrow\right>}
\newcommand{\updown}     {\left|\uparrow\downarrow\right>}
\newcommand{\upup}          {\left|\uparrow\uparrow\right>}

\begin{document}
\title{Spectroscopy on two coupled flux qubits}
\author{J.~B. Majer$^1$, F. G. Paauw, A. C. J. ter Haar, C. J. P. M. Harmans, and J.~E. Mooij}
\date{\today}
\affiliation{  Department of Nanoscience, Delft University of
Technology,
 Lorentzweg 1, 2628 CJ Delft, The Netherlands\\ 
 $^{1}$Present Address: Department of Applied Physics, Yale University, New Haven, CT 06511\\ }
\begin{abstract}
We have performed spectroscopy measurements on two coupled flux qubits. The qubits are coupled inductively, which results in a $\sigma_1^z\sigma_2^z$ interaction. By applying microwave radiation, we observe resonances due to transitions from the ground state to the first two excited states. From the position of these resonances as a function of the magnetic field applied we observe the coupling of the qubits. The coupling strength agrees well  with calculations of the mutual inductance.
\end{abstract}
\maketitle
Quantum computers manipulate quantum information contained in the states of interacting two-level systems called quantum bits or qubits. Quantum gates require qubit-qubit coupling that preserves quantum coherence \cite{NielsenChuang}. Qubits have been implemented in various systems. However, the requirement for upscaling makes solid state implementations highly attractive. Several single solid state qubits have been realized using superconducting Josephson junction circuits \cite{Mooij99,Nakamura99,Friedman00,Vion02,Han02,Martinis02}. 
As a next step coupled multiple qubits need to be studied. Recently charge quantum dynamics in two coupled Josephson qubits has been observed \cite{Pashkin03}. Spectroscopic measurements on coupled charge qubits have been performed as well \cite{Berkley03}. 

In this letter we present spectroscopy measurements on two coupled flux qubits \cite{Mooij99}.  Our flux qubit consists of a small superconducting loop interrupted by three Josephson junctions with a small junction capacitance $C$. When approximately half a flux quantum $\Phi_0$ is applied to the ring the two classical states correspond to clockwise  and anti-clockwise  circulating currents. The circulating current of the  qubit generates a magnetic field into $\down$ or out of $\up$ plane.
Due to the large charging energy $E_C=e^2/2C$ the barrier between these classical states is low and tunneling between the states will take place. When exactly half a flux quantum is applied, the two eigenstates are symmetric and antisymmetric superpositions of the two classical states. 
The system is described by the effective Hamiltonian
$\mathcal{H}=h\sigma^z+t\sigma^x$ where $\sigma^{z,x}$ are Pauli spin matrices and  $h=I_p(\Phi-1/2 \Phi_0)$, with $I_p$ the circulating current and $\Phi$ the flux applied to the qubit. The tunneling amplitude $t$ depends exponentially on the ratio $E_J/E_C$ \cite{Orlando99}. Here $E_J=I_0 \Phi_0/2\pi$ is the Josephson energy, with $I_0$ the critical current of one junction.
Because the basic states of this qubit are flux states it is very insensitive to charge noise.
The energy level repulsion due to the tunneling amplitude $t$ was observed with spectroscopy\cite{VanderWal00}. Recently also coherent oscillations of such a flux qubit have been observed \cite{Chiorescu03}.

\begin{figure}
\includegraphics[width=7cm]{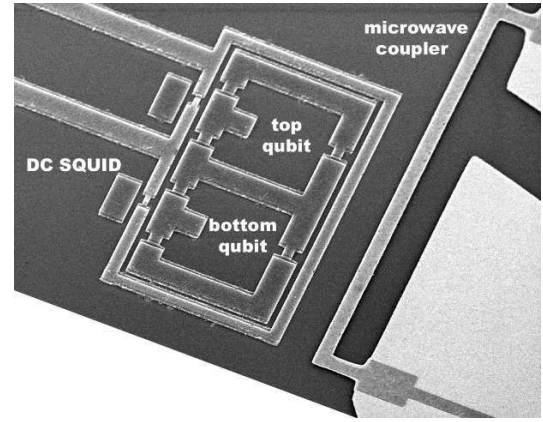}
\caption{
Scanning microscope image of the two coupled qubits surrounded by the DC-SQUID. A part of the microwave coupler is visible on the right.  The width of the qubits is 5 $\mu$m.}
\label{SEM}
\end{figure}

The easiest way to realize inter-qubit coupling for flux qubits is by magnetic interaction. This coupling results from the magnetic flux that the circulating current in one qubit induces in the second qubit. The interaction is of Ising form $\sigma_1^z\sigma_2^z$, since the clockwise $\down$ and the anti-clockwise $\up$ circulating states give a z-component field. The coupling strength $j$ is given by
\begin{equation}
j=2 M I_{p,1}I_{p,2}
\label{couplingstrength}
\end{equation}
where $M$ is the mutual inductance between the qubits.
$M I_{p,1}$ is the generated flux in qubit 2 by the current in qubit 1 leading to an increase in
energy of $M I_{p,1} I_{p,2}$ in the second qubit.
The sign of the coupling is such that it favors anti-parallel qubit states. In the two level description the self induced flux can be ignored because it adds a constant energy term.
Figure \ref{SEM} shows a scanning electron microscope image of the two qubits. In order to enhance the qubit-qubit mutual inductance $M$ the two qubits share one branch. 

The two qubits are now described by the total Hamiltonian:
\begin{eqnarray}
&\mathcal{H}
=h_1\sigma_1^z+t_1\sigma_1^x
+h_2\sigma_2^z+t_2\sigma_2^x
+j\sigma_1^z\sigma_2^z&\nonumber\\
&={\small\left(
\begin{array}{cccc}
-h_1\!-\!h_2\!+\!j\!\! & t_2 & t_1 & 0\\
t_2 & \!\!-h_1\!+\!h_2\!-\!j\!\! & 0& t_1 \\
t_1 & 0 & \!\!h_1\!-\!h_2\!-\!j\!\! & t_2 \\
0 & t_1&t_2&\!\!h_1\!+\!h_2\!+\!j
\label{Hamiltonian}
\end{array}
\right)}&
\end{eqnarray}

Here $h_1=I_{p,1}(\Phi_1-1/2 \Phi_0)$  and $t_1$ is the tunneling amplitude of the first qubit. Similar expressions hold for the second qubit.  

\begin{figure}
\includegraphics[width=7cm]{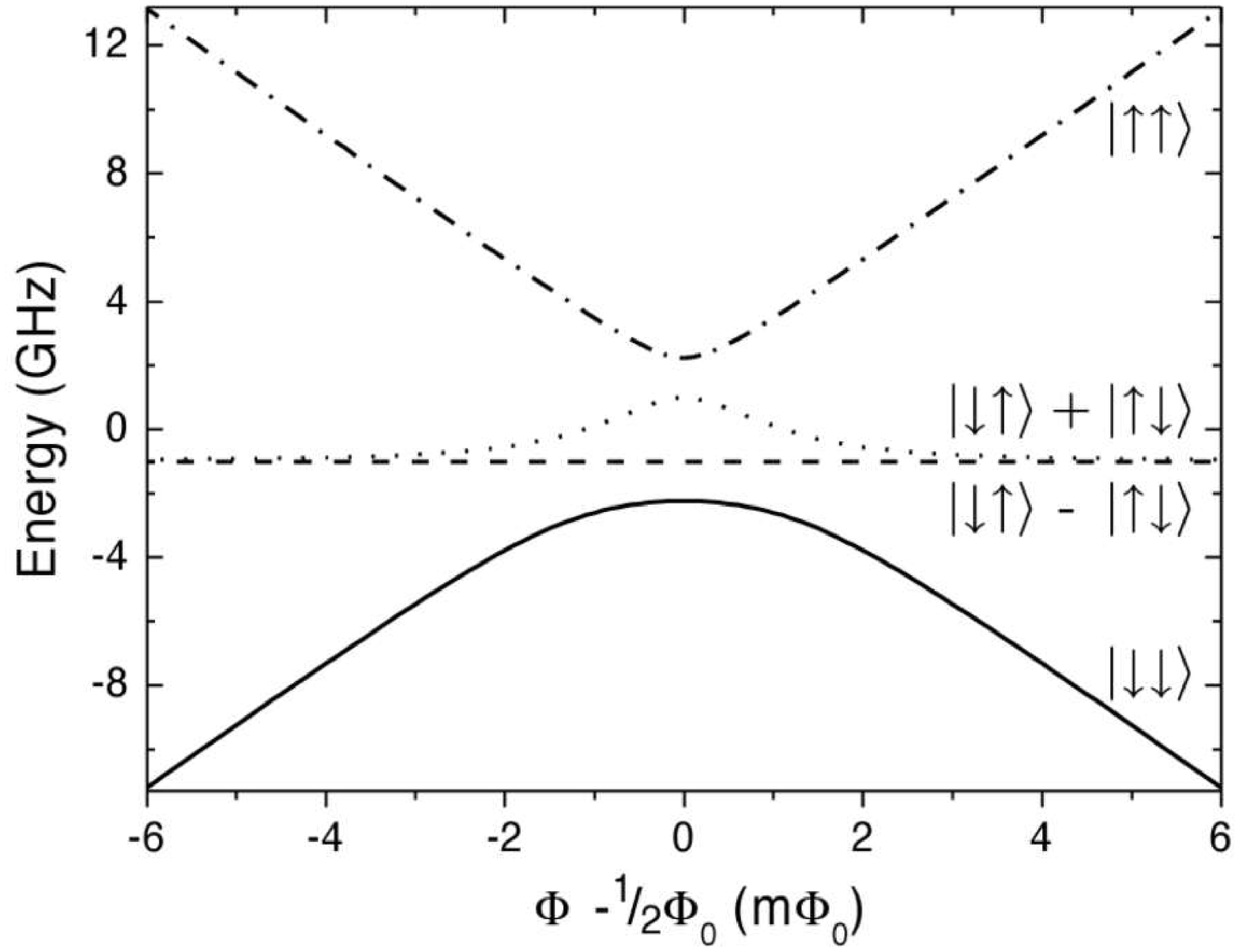}
\caption{Energy levels of two identical coupled qubits as a function of the externally applied flux. Far away from the degeneracy point $|h|\gg t$, the energy states resemble classical states.  The coupling $j$ shifts the energy of the ferromagnetic states $\upup$, $\downdown$ up, while the energy of the antiferromagnetic states $\updown$, $\downup$ is lowered. In the vicinity of the degeneracy point $|h|\lesssim t$, the classical states mix due to the tunnel coupling $t$. Due to the symmetry of the problem, the anti-symmetric energy state $\downup-\updown$ does not mix with symmetric states.
}
\label{Energy_sym}
\end{figure}

The energy levels and energy eigenstates of two coupled qubits can be calculated by diagonalizing the Hamiltonian (\ref{Hamiltonian}). A discussion for coupled solid-state qubits is given in \cite{Storcz03}. Figure \ref{Energy_sym} shows the energy levels for two identical qubits. The parameters are $I_{p,1}=I_{p,2}=$ 1 GHz / m$\Phi_0$ and $j=t_1=t_2$= 1 GHz. Far away from the degeneracy point ($|h|\gg t$), the energy states closely resemble  the  classical states $\downdown$, $\downup$, $\updown$, $\upup$. The two antiferromagnetic states $\updown$, $\downup$ (with opposite circulating currents in the qubits) are degenerate. The coupling $j$ reduces the energy of the antiferromagnetic states and increases the energy of the ferromagnetic states $\upup$, $\downdown$. In the vicinity of the degeneracy point ($|h|\lesssim t$), the states are superpositions of the classical states due to the tunnel coupling $t$. Due to the symmetry of the problem the anti-symmetric energy state $\downup-\updown$ does not mix with the other states and its energy is independent of the magnetic field applied. 

\begin{figure}
\includegraphics[width=7cm]{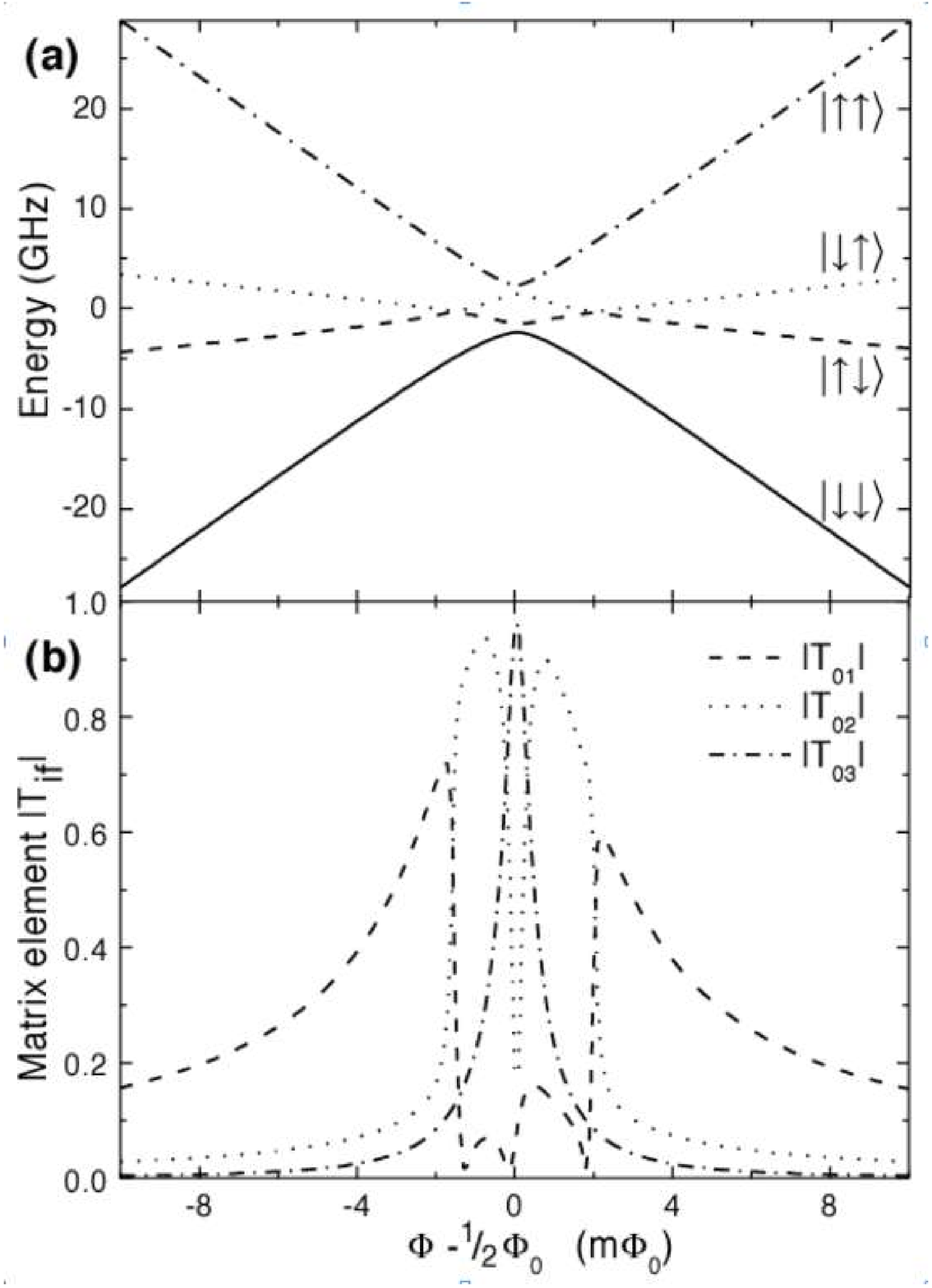}
\caption{(a) Energy levels of two different coupled qubits as a function of the externally applied flux. Due to the difference between the two qubits, the degeneracy of the two antiferromagnetic states $\updown$, $\downup$ is lifted \\(b) Transition matrix elements $|T_{if}|=|\left<\Psi_f\right|\sigma_1^z+\sigma_2^z\left|\Psi_i\right>|$ for the transitions from the ground state to the three excited states. The matrix element for the transition from the ground state to the third excited state is very small except for the vicinity of the degeneracy point. The parameter values are taken from the values obtained by fitting the experimental data (Fig. \ref{Spectroscopy}).
}
\label{EnergyTransition}
\end{figure}

In practice the two qubits have different parameter values due to fabrication limitations. Figure \ref{EnergyTransition}(a) shows the energy levels for two unequal qubits. The parameter values are those of the actual sample (as obtained below).
Because the qubits have different circulating currents $I_{p,1}\neq I_{p,2}$ the degeneracy of the antiferromagnetic states is lifted. 

The signal of the two qubits is measured with a DC superconducting quantum interference device (SQUID). The SQUID surrounds the two qubits (Fig. \ref{SEM}) and measures the magnetization, i.e. generated flux, of the two qubits together. The magnetization of the two qubits is given by the slope of the energy levels as a function of the magnetic flux (Fig. \ref{Energy_sym}, \ref{EnergyTransition}). 
On the right-hand side of figure \ref{SEM}(a) part of the microwave coupler is visible. When the frequency of the microwave matches the energy difference between two energy states of the coupled qubits the radiation induces transitions between the two states. The microwave coupler induces approximately the same amount of flux in the two qubits. Therefore the transition matrix element is of the form $\sigma_1^z+\sigma_2^z$. Figure \ref{EnergyTransition}(b) shows the absolute value of the transition elements $T_{if}=\left<\Psi_f\right|\sigma_1^z+\sigma_2^z\left|\Psi_i\right>$

\begin{figure}
\includegraphics[width=8.5cm]{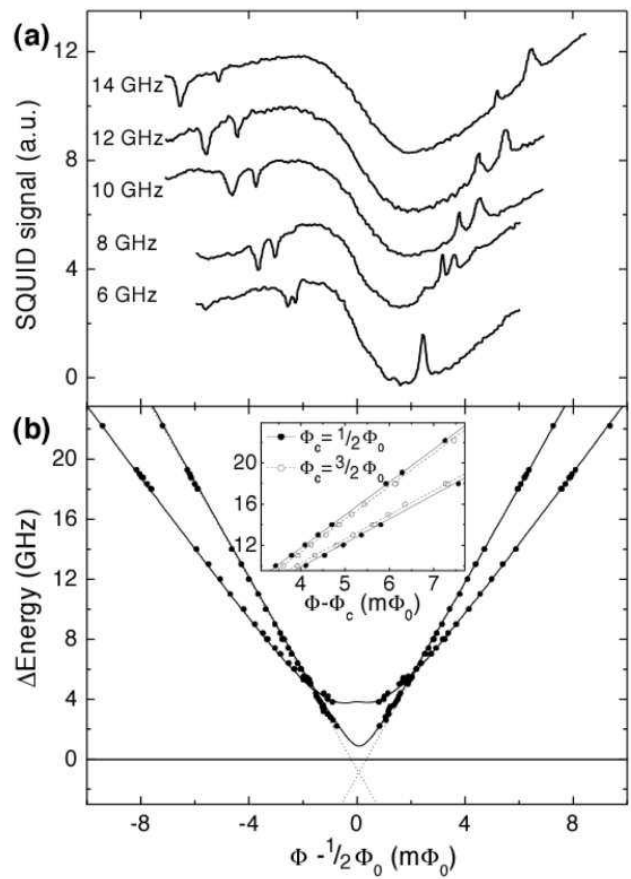}
\caption{Spectroscopy measurements. (a) SQUID signal, i.e. magnetization of the two qubits, as a function of the applied magnetic field. The microwaves induce transition from the ground state to the first and second excited state, which appear as dips and peaks in the magnetization. The outer peaks and dips are due to the first excited state and the inner due to the second.
(b) Microwave frequency versus peak and dip position. In the vicinity of the degeneracy point ($\Phi\approx 1/2\Phi_0$), the resonances deviate from straight lines because the energy levels are formed by superposition of the classical states. 
The straight line is a fit of the two qubit Hamiltonian (Eq. \ref{Hamiltonian}) to the data. 
The inset compares the data obtained in the vicinity of $3/2\Phi_0$ with the data obtained at $1/2\Phi_0$. The shift is due to the influence of the different qubit areas which is three times higher. 
}
\label{Spectroscopy}
\end{figure}

Measurements were performed in a dilution refrigerator at a base temperature of 20 mK. We measured the switching current of the DC-SQUID. This was done by ramping the bias current and recording the current, where the SQUID switched to a voltage state \cite{VanderWal00}.  We repeated this procedure and averaged the switching current over typically 1000 measurements. We varied the field in a 20 m$\Phi_0$ range centered at 1/2 $\Phi_0$. The SQUID signal shows a characteristic step (Fig.  \ref{Spectroscopy}(a)). The ground state of the two qubits changes from a $\upup$ state via a superposition of all states at 1/2 $\Phi$ to the $\downdown$ state. Applying  microwaves leads to dips on the left of the step and peaks on the right of the step (Fig. \ref{Spectroscopy}(a)). At these positions the energy of the microwaves matches the difference between energy levels and induces transitions between them. 
As the system of the two qubits is excited continuously an incoherent
mixture is formed and the SQUID signal is the average of the magnetization of the two levels.
We have studied the positions of these resonances as a function of the applied microwave frequency. Figure \ref{Spectroscopy}(b) shows the frequency of the resonance versus the position. Far away from 1/2 $\Phi_0$  the peaks and dips approximately follow straight lines. Here the resonances correspond to transitions between states where one of the two qubits is flipped. One observes that the slopes of the two resonances and consequently the persistent currents of the two qubits are different.
The dotted lines in \ref{Spectroscopy}.b are linear fits to the data above 8 GHz for one of the qubits. From these lines an estimate can be made of the coupling strength $j$. As the ground state is ferromagnetic and has its levels shifted up by $j$, while the excited state is antiferromagnetic with levels shifted down by j, the intersection should occur at a level of $-2j$. For the dotted lines one finds a crossing at a level of -0.98 GHz, indicating a value $j$~=~0.49~GHz. The transitions for the other qubit yield a less accurate estimate as can be understood later from the full parameter fit (the higher value of $t_2$ leads to strong rounding).
One observes that the resonance peaks and dips cross at two points at opposite sides of the degeneracy point 1/2 $\Phi_0$ with the crossing points at $\sim$ -1.5 m$\Phi_0$ for the peaks and $\sim$ +2.2 m$\Phi_0$ for the dips. This is a peculiar feature of the quantum nature of the two qubits. If the qubits were classical ($t_{1}=t_{2}=0$) the energy levels in figure \ref{EnergyTransition}(a) would follow straight lines. The first and second excited state lines thus could cross at one point only.

In order to obtain all coupled qubit parameters we did a least-squares fit to the difference of energy levels calculated with the coupled qubit Hamiltonian (\ref{Hamiltonian}). 
Here we make the assumption that the magnetic field is homogeneous for the qubits and the SQUID. The least square fit involves six fitting parameters: the persistent currents $I_{p,1}$ and $I_{p,2}$, the tunneling amplitudes $t_1$ and $t_2$, the inter-qubit coupling strength $j$, and the relative difference in area of the two qubits $\sigma$.
From the fit we obtain the following parameters for the two qubits: $I_{p,1}$~=~512nA~$\pm$~6nA, $t_1$~=~0.45~GHz~$\pm$~0.2~GHz and $I_{p,2}$~=~392~nA~$\pm$~5nA, $t_2$~=~1.9~GHz~$\pm$~0.1~GHz. The persistent current is directly dependent on the critical current of the Josephson junctions \cite{Orlando99}. The difference in persistent currents of the two qubits is due to the spread in critical currents in the fabrication process. While the tunneling amplitude $t$ depends exponentially on $E_J/E_C$, the typical spread in the junction parameters  can easily lead to a difference as large as observed. The qubit coupling strength obtained from the fit is  
$j$~=~0.50~GHz~$\pm$~0.03~GHz and agrees well with result from the analysis above. The difference in qubit area is $\sigma$~=~0.027\%~$\pm$~0.004\%. If the areas would differ more than a few parts in a thousand, the degeneracy points of the two qubits would be so far apart that one would measure a double step. 

The coupling strength $j$ can also be evaluated using Eq. (\ref{couplingstrength}). From the geometry the mutual inductance can be calculated, yielding $M$~=0.86~pH. Using the values persistent currents obtained in the fitting procedure, we get a coupling strength of $j=0.52$ GHz, which agrees well with the result above.

We have also performed the spectroscopy measurements in the vicinity of 3/2 $\Phi_0$. 
At this point everything is equivalent to the situation at 1/2 $\Phi_0$ except for the flux bias due to the difference of the areas. The effect of the different qubit areas on the energy levels is now three times higher. The inset in figure \ref{Spectroscopy}(b) compares the data at 3/2 and 1/2 $\Phi_0$ at high frequencies. The peaks due to the first excited state are shifted to lower magnetic field values and the peaks of the second excited state to higher values. From the magnitude of the shift one can calculate the difference of the qubit areas to be $\sigma$~=~0.03\%, which agrees with the result from the fit above.

We have only observed transitions from the ground state to the first two excited states, but not to the third excited state. Figure \ref{EnergyTransition}(b) shows that the transition amplitude $T_{03}$ for this resonance is very small except for the vicinity of the degeneracy point 1/2 $\Phi_0$. However, near the degeneracy point the ground state and the third excited state have similar slope and therefore similar magnetization. Consequently we are not able to observe the transition with the SQUID. Figure \ref{EnergyTransition}(b) also shows that the transition amplitude to the first excited state is larger than to the second excited state. In agreement with that prediction, one observes that the outer resonance peaks and dips are broader than the inner ones  (Fig. \ref{Spectroscopy}(a)).

The previous analysis makes clear that the spectroscopic data are fully consistent with the two-qubit Hamiltonian of Eq.(\ref{Hamiltonian}). This Hamiltonian, in turn, opens the possibility of well-chosen one and two-qubit operations that lead to controlled entanglement. The new results support the notion that superconducting flux qubits can be used to study entanglement in macroscopic quantum systems and for the development of non-trivial two-qubit gates such as the controlled-not.

In summary, we have performed spectroscopy measurements on two coupled flux qubits. The mutual inductance between the two qubits leads to Ising coupling $\sigma_1^z\sigma_2^z$. The observed resonances agree very well with the two qubit Hamiltonian (Eq. (\ref{Hamiltonian})). The magnitude of the qubit coupling agrees well with the calculation of the mutual inductance between the qubits. 

We thank T. Orlando, L. Levitov and M. Devoret for discussions as well as R. Schouten for technical assistance. This work was supported by the Dutch Foundation for Fundamental Reseach on Matter (FOM), the European Union SQUBIT project, and the U.S. Army Research Office (grant DAAD 19-00-1-0548).

\end{document}